\documentclass{anstrans}
\title{One-Way Coupled Tumor Response Model for Combined-Hyperthermia-Radiotherapy Treatment \\ with Anisotropic Scattering}
\author{Japan K.~Patel, John J.~Kuczek, and Richard Vasques}

\institute{
Department of Mechanical and Aerospace Engineering, The Ohio State University, \\
201 W. $19^{th}$ Avenue, Columbus, OH, 43210 \\ patel.3545@osu.edu, kuczek.6@buckeyemail.osu.edu, vasques.4@osu.edu
}

\usepackage{graphicx} 
\usepackage{booktabs} 
\usepackage{microtype} 
\usepackage[capitalize]{cleveref}
\usepackage{amsbsy}
\usepackage{amsmath}
\usepackage{gensymb}
\usepackage{float}
\usepackage{graphicx}

\begin{document}
\section{INTRODUCTION}
\vspace{-5pt}

Radiotherapy is one of the most widely used cancer treatment modalities \cite{radbio}. 
Its primary goal is to shrink tumors and kill cancer cells without permanently damaging the surrounding healthy tissue. 
This is managed by carefully manipulating dose profiles and delivery schedules, and by boosting cancer cell death using radiosensitizers \cite{radonc}. 
Studies have confirmed that tumor heating changes several critical parameters within the tumor microenvironment. 
Specifically, hyperthermia enhances vascular perfusion and oxygenation state \cite{radbio}. 
Moreover, it inhibits the repair of DNA \cite{radbio}.
Hyperthermia is, therefore, an excellent radiosensitizer and makes tumors more susceptible to death through radiation \cite{hyperrt}.
Therapies such as combined-hyperthermia-radiotherapy (CHR) make use of this, requiring treatment of tumors with heat and radiation.
Several phase III clinical trials for cervical cancer, superficial breast cancer, malignant metastatic melanoma, head and neck cancer, and glioma have demonstrated significant benefits of using hyperthermia in combination with radiotherapy \cite{radbio}.

Mathematical tumor response modeling is increasingly being recognized as one of the most effective tools to provide insights on dose administration \cite{radmichor}.
While there are several single-physics models describing radiotherapy and tumor response \cite{altrock,mmtumgro,sachs,rockne}, few models consider the interdependence of the physics involved.
To appropriately model a CHR treatment, features like tumor heating (heat transfer), dosimetry (radiation transport), and tumor dynamics (cell population dynamics) need to be considered together. 
Our first paper on modeling such treatments introduced a one-way coupled model that could only account for isotropic scattering \cite{chr1}. 
In the present work, we extend our transport model to incorporate anisotropic scattering. 
We note that this is not an exhaustive study; in particular, the test problem discussed here \textit{does not} model a real-world treatment scenario.

The remainder of this paper is organized as follows. 
The governing equations and relevant physics are introduced in the next section. 
Subsequently, we describe a test problem and present numerical results. 
We discuss our conclusions in the final section.
\vspace{-5pt}

\section{GOVERNING EQUATIONS}
\vspace{-5pt}

Our CHR model is driven by three main physics: 1) radiation transport to model internal dosimetry, 2) heat transfer to incorporate the effects of tumor heating, and 3) cell population dynamics to determine how the tumor responds to the combination treatment. 
In this work, we consider a simple model that is is one-dimensional in space, monoenergetic, and accomodates anisotropic radiation sources and scattering.
\vspace{-5pt}

\subsection{Radiation Transport and Dosimetry}
\vspace{-5pt}

The standard Boltzmann transport equation \cite{lm} adequately models transport in problems with forward-peaked scattering.
However, sometimes, prohibitively large Legendre expansion orders are required to accurately represent anisotropic kernels \cite{aydin}.
In order to overcome this limitation, the scattering term is broken into smooth and singular components \cite{landesman}.
Such decomposition of the transport equation leads to the Boltzmann-Fokker-Planck (BFP) approximation. 
We use the following monoenergetic Boltzmann-Fokker-Planck equation to model the flux distribution \cite{landesman}:
\begin{subequations}
\begin{equation}
\begin{split}
\label{transport1}
\frac{1}{v}\frac{\partial \psi}{\partial t} + \mu\frac{\partial \psi}{\partial x} + \sigma_t(x) \psi =  \sum_{l=0}^{L-2} \frac{2l+1}{2} P_l(\mu)\tilde{\sigma}_{s,l} \phi_l \\ + \frac{\sigma_{tr}}{2} \frac{\partial}{\partial \mu} (1-\mu^2) \frac{\partial \psi}{\partial \mu} + Q(x, \mu, t),
\end{split}
\end{equation}
\begin{equation}
\label{bcl}
\psi(0. \mu, t) = a(\mu, t) \quad \text{for} \quad \mu > 0,
\end{equation}
\begin{equation}
\label{bcr}
\psi(X, \mu, t) = b(\mu, t) \quad \text{for} \quad \mu < 0, 
\end{equation}
\end{subequations}
where $\psi = \psi(x,\mu, t)$ represents the angular flux in terms of position $x$, direction $\mu$, and time $t$; $v$ is the photon velocity; $\phi$ is the scalar flux; $\sigma_t$ is the total macroscopic cross-section; $L$ is the Legendre expansion order for scattering; $a$ and $b$ are the prescribed boundary conditions; and $Q$ is an anisotropic source.
Moreover, $\tilde{\sigma}_{s, l}$ represents the $l^{th}$ moment of the smooth component of the scattering cross-section $\sigma_s$.
The Legendre moments of the scattering cross-section are represented by $\sigma_{s,l}$.
The moments smooth component of the scattering cross-section is written as \cite{landesman}:
\begin{equation}
\begin{split}
\label{smooth}
\tilde{\sigma}_{s,l} = \sigma_{s,l} - \sigma_{s, L} =\frac{\sigma_{tr}}{2}\left[L(L+1) - l(l+1) \right], \\ l=0, 1...L-2.
\end{split}
\end{equation}
The momentum transfer $\sigma_{tr}$ is represented as:
\begin{equation}
\label{momentum_transfer}
\sigma_{tr} = \frac{\sigma_{s,L-1} - \sigma_{s, L}}{L}.
\end{equation}
We use the following Henyey-Greenstein kernel to represent the anisotropic scattering:
\begin{equation}
\label{hgk}
f(\mu) = \frac{1 - g^2}{2(1 + g^2 -2g\mu)^{3/2}},
\end{equation}
where $f$ is the phase function used to represent scattering and $g$ is the anisotropy factor. The moments of this phase function are \cite{hgk}:
\begin{equation}
\label{moments_fl}
f_l = g^l.
\end{equation}
We calculate the dose using the following equations \cite{martin}:
\begin{subequations}
\begin{equation}
\label{dose1}
D(x, t) = \left(\frac{\mu_{en}}{\rho}\right)\Phi(x, t) E,
\end{equation}
\begin{equation}
\label{fluence}
\Phi(x, t) = \int_{0}^{t} dt' \phi(x, t').
\end{equation}
\end{subequations}
Here, $D$ represents the dose; $\frac{\mu_{en}}{\rho}$ is the mass energy absorption coefficient; $E$ is the energy per photon; and $\Phi$ is the fluence at time $t$. 
The effective dose $D_E$ is evaluated using the linear-quadratic model \cite{radbio}:
\begin{equation}
\label{edose1}
D_E(x, t) = \alpha D(x, t) + \beta D^2(x,t),
\end{equation}
where $\alpha$ and $\beta$ are radiobiology parameters that determine relative contribution of each term in the sum toward the total radiation effect \cite{radbio}. 
\vspace{-5pt}

\subsection{Hyperthermia and Heat Transfer}
\vspace{-5pt}

Recently, there has been an increase in the use of thermal medicine for cancer care. 
Novel techniques, including radiofrequency and ultrasound ablation, have been developed in order to induce hyperthermia inside tumors \cite{radbio}.
Although the use of a more sophisticated heat transport model is warranted, here we use simple heat conduction to model hyperthermia \cite{cannon}:
\begin{equation}
\label{heat}
\rho c_p \frac{\partial T(x, t)}{\partial t} = \frac{\partial }{\partial x} \kappa \frac{\partial T(x, t)}{\partial x} + q(x),
\end{equation}
where $c_p$, $\rho$, and $\kappa$ are respectively the specific heat capacity, the material density, and the thermal conductivity of tissue (assumed constant), and $q$ is the volumetric heat source.
\vspace{-5pt}

\subsection{Cell Survival and Radiosensitivity Parameter}
\vspace{-5pt}

The cell survival probability $S$ is chosen in such a way that a larger dose results in smaller survival probability, as given by \cite{rockne}:
\begin{equation}
\label{survival}
S(x,t) = e^{-D_E(x, t)}.
\end{equation}

In order to incorporate heat effects, we use the radiosensitivity parameter \cite{chr1} $\xi$. 
We define this parameter as the ratio of the biological damage caused by a given amount of radiation dose to tissue with and without heating.
We represent this damage (cell kill) as $R_{CHR}$ and $R_{RT}$ respectively: 
\begin{subequations}
\begin{equation}
\label{rrt}
R_{RT} = 1 - S,
\end{equation}
\begin{equation}
\label{rchr}
R_{CHR} = \xi(1-S).
\end{equation}
\end{subequations}
At normal body temperature, $\xi =1$ irrespective of time. 
We also set $\xi$ at time zero to be unity.
We arbitrarily assume $\xi=2.5$ for our system when the tumor is heated to an average temperature $T_{avg}$ of $45 \degree$C over $30$ minutes.
In order to define $\xi$ elsewhere, we employ bilinear interpolation \cite{nm} between the data points $\xi(37, 0) = 1$, $\xi(37, 30)=1$, $\xi(45, 0)=1$  and $\xi(45, 30)=2.5$. This returns:
\begin{equation}
\label{rsp}
\xi(T_{avg},t) = 1 - 0.23125T_{avg} + 0.00625(T_{avg}\times t).
\end{equation}
\vspace{-5pt}

\subsection{Cell Population Dynamics}
\vspace{-5pt}

Traditionally, the tumor dynamics are determined according to three factors: proliferation, invasion, and cell kill induced by treatment \cite{rockne}. 
Proliferation determines the balance between cell division and cell loss due to natural death; 
invasion determines the transport of tumor cells in the area of interest; and the cell kill incorporates the effects of treatment on tumors.
The following balance equation determines tumor cell concentration, $c(x,t)$, over time \cite{rockne}:
\begin{equation}
\label{td1}
\frac{\partial c}{\partial t} = \frac{\partial}{\partial x}\left(I\frac{\partial c}{\partial x}\right) + pc(x,t) - R(x, t)c(x,t).
\end{equation}
Here, the diffusion coefficient $I$ represents motility of tumor cells, and $p$ is the proliferation rate.
Patient-specific invasion and proliferation rates can be obtained using MRI images \cite{rockne}.

This preliminary model is heavily simplified and does not currently consider temperature-dependent cross-sections for transport, material density changes, and decay heat.
These simplifications allow us to get a one-way coupled system.
We plan to eliminate these assumptions and model this system with a fully-coupled framework in a later paper.
\vspace{-5pt}

\subsection{CHR Model}
\vspace{-5pt}

The tumor response over time is determined using the equations presented in the previous section. The coupling and solution can be represented via the following Newton step:
\begin{equation}
\left[\begin{array}{ccccc}
J_{\phi \phi} & 0 & 0 & 0 & 0 \\
J_{\phi D_E} & J_{D_E D_E} & 0 & 0 & 0 \\
0 & 0 & J_{T T} & 0 & 0 \\
0 & J_{D_E R} & J_{T R} & J_{R R} & 0\\
0 & 0 & 0 & J_{R c} & J_{c c}
\end{array}\right] \left[\begin{array}{c}
\delta \phi \\
\delta D_E \\
\delta T \\
\delta R \\
\delta c
\end{array}\right] = -\left[\begin{array}{c}
F_{\phi} \\
F_{D_E} \\
F_T \\
F_R \\
F_c
\end{array} \right],
\end{equation} 
where $F_{\phi}$, $F_{D_E}$, $F_T$, $F_R$, and $F_c$ are the residual forms of the relevant equations for flux, effective dose, temperature, effect of therapy on cell kill, and tumor cell concentration. 

Since the Jacobian matrix above is not block-diagonal, the physics are interrelated. 
The lower-triangular structure of the Jacobian matrix suggests one-way coupling.
Therefore, we solve the problem in a serial fashion.
First, we solve the BFP equation to determine flux.
Subsequently, fluence, dose, effective dose, and the survival probability (in that order) are evaluated.
Next, we solve the heat equation to determine the temperature, and then evaluate the heating-adjusted effect of radiation on tissue.
Finally, the tumor cell concentration is evaluated.
This kind of modeling allows us to continuously track tumor cell density over time.

We discretize the BFP equation using the standard backward Euler method in time and diamond-difference/discrete ordinates scheme in space/angle.
The angular Laplacian uses Morel's weighted finite difference scheme \cite{morel_fp}.
The heat transfer and the cell population dynamics equations employ backward Euler and central finite difference for time/space discretization.
These schemes are well-known and therefore are not discussed here in detail. 
\vspace{-5pt}

\section{NUMERICAL RESULTS AND DISCUSSION}
\vspace{-5pt}

We consider the evolution of tumor-cell concentration distribution in a three-region slab over a period of thirty minutes. 
The first and the third regions, each $1.4$ cm thick, represent healthy soft-tissue.
The second soft-tissue region (in the middle) is $0.2$ cm thick, and has uniformly distributed tumor cells with a concentration of $2^{20}$ $\frac{cells}{cm^3}$.
Moreover, in order to induce hyperthermia, this region also has a constant volumetric heat source.
We introduce radiation to this system through a time-dependent beam of $0.4$ MeV photons throught the left boundary along the most forward discrete direction $\mu_N$; that is:
\begin{equation}
\label{a}
\psi(0, \mu, t)  = a(\mu, t)  = \left\{
\begin{array}{ll}
5.6\times 10^5 e^{-\eta t}, &\text{if} \quad \mu = \mu_N\,, \\
0, &\text{otherwise}\,.
\end{array}
\right.
\end{equation}
Here, $\eta$ is 75 days.
We also assume a vacuum right boundary such that $\psi(X.\mu,t) = b(\mu,t) = 0$, with $X=3$ cm.

The heat source is $q = \frac{1}{3}$ $\frac{W}{cm^3}$.
Both the heat transfer and tumor dynamics equations have open boundaries. 
We assume the parameters presented in \cref{tab1} \cite{rockne,warrell}.
\begin{table*}[htb]
  \centering
  \caption{\bf{Parameters}}
\begin{tabular}{ |c|c|c|c|c|c|c|c|c|}\hline\label{tab1}
 \textbf{I}  & \textbf{p} & \pmb{$\alpha$}  & \pmb{$\frac{\alpha}{\beta}$}  & $\bm\kappa$ & \pmb{$c_p$} & \pmb{$\rho$}  & \pmb{$\frac{\mu_{en}}{\rho}$} \\
  $\mathrm{[mm/yr]}$ & [/yr]  & [kg/J]&[$\mathrm{J/kg}$] &[W/m.K] &[kJ/kg.K] &[g/cc] & [$\mathrm{cm^2/g}$]  \\
  \hline
  4.29 & 35.13 & 0.203 & 10 & 0.51 & 3.68 & 1.0 & 0.0325\\ \hline 
\end{tabular}
\end{table*}
We follow existing literature and assume similar continuum properties for both healthy and tumor tissue \cite{rockne}.
The relevant cross-section moments are presented in \cref{tab2}.
\begin{table*}[htb]
  \centering
  \caption{\bf{Scattering Cross-section Moments}}
\begin{tabular}{c c || c c} \hline \label{tab2}
Moment & Value & Moment & Value \\ \hline \hline
 $\sigma_{t}$ & 0.1053578  & $\sigma_{tr}$ & 0.000333856436591749  \\
 $\sigma_{s,0}$ & 0.105324 & $\tilde{\sigma}_{s,0}$ & 0.050237687962361  \\
 $\sigma_{s,1}$ & 0.0947916 & $\tilde{\sigma}_{s,1}$ & 0.040039144398953  \\
 $\sigma_{s,2}$ & 0.08531244 & $\tilde{\sigma}_{s,2}$ & 0.031227697272137  \\
 $\sigma_{s,3}$ & 0.076781196 & $\tilde{\sigma}_{s,3}$ & 0.023698022581912  \\
 $\sigma_{s,4}$ & 0.0691030764 & $\tilde{\sigma}_{s,4}$ & 0.017355328728279 \\
 $\sigma_{s,5}$ & 0.06219276876 & $\tilde{\sigma}_{s,5}$ & 0.012114303271238 \\
 $\sigma_{s,6}$ & 0.055973491884 & $\tilde{\sigma}_{s,6}$& 0.007898165014788  \\
 $\sigma_{s,7}$ & 0.0503761426956 & $\tilde{\sigma}_{s,7}$ & 0.004637810882530  \\ 
 $\sigma_{s,8}$ & 0.04533852842604 & $\tilde{\sigma}_{s,8}$ & 0.002271048105704  \\
 $\sigma_{s,9}$ & 0.040804675583436 & $\tilde{\sigma}_{s,9}$ & 0.000741903192426  \\
 $\sigma_{s,10}$ & 0.036724208025092 & $\tilde{\sigma}_{s,10}$ & 0  \\
 $\sigma_{s,11}$ & 0.033051787222583 & $\tilde{\sigma}_{s,11}$ & 0  \\ \hline
\end{tabular}
\end{table*}
For each of the test problems, we choose a uniform time step of $\Delta t = 0.025$ $s$ and $270$ spatial cells.
We set the angular discretization for the transport equation to twelve angles.
We assume an anisotropy factor of $g = 0.9$. 
The BFP equation is solved using the standard source iteration method, and the heat and tumor dynamics equations are solved using MATLAB's backslash function \cite{matlab}.
\vspace{-5pt}

\subsection{Test Problem}
\vspace{-5pt}

\begin{figure}[ht] 
  \centering
  \includegraphics[scale=0.45]{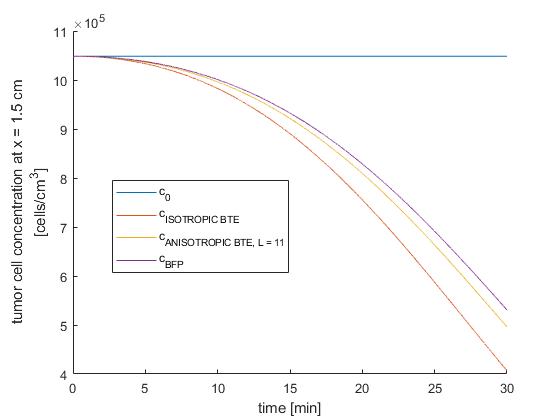}
  \caption{Tumor Cell Concentration Comparison}
  \label{fig:fig1}
\end{figure}

First, we analyze the estimates for the final tumor cell concentration when using different models for transport (with a highly anisotropic beam source).
We compare three transport models: 1) transport equation with isotropic scattering; 2) transport equation with anisotropic scattering with a truncated scattering expansion order; and 3) BFP equation.
\Cref{fig:fig1} plots tumor cell concentration at the center-most node of the spatial grid over time. We note that both isotropic and anisotropic transport with a truncated scattering term underestimate tumor cell population over time. This can be attributed to the fact that both these models do not represent scattering adequately.
Therefore, these models do not accurately model (overestimate) flux evolution over time.
This results in an underestimation of tumor cell concentration.

Next, we consider tumor dynamics in three scenarios: 1) untreated tissue (no radiation or heat source), 2) tumor undergoing stand-alone radiotherapy, and 3) tumor combined-hyperthermia. We maintain the same heat and radiation sources in our system for the relevant physics. We model transport using the BFP equation.
\begin{figure}[ht] 
  \centering
  \includegraphics[scale=0.45]{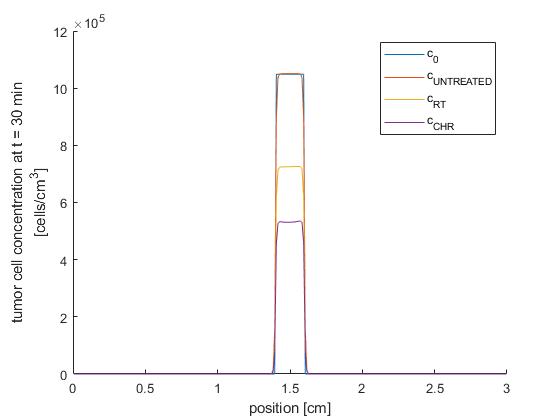}
  \caption{Treatment Comparison}
  \label{fig:fig2}
\end{figure}
\Cref{fig:fig2} shows that the radiation beam begins to kill cancer cells and reduces the total tumor cell concentration at the end of thirty minutes. 
We also observe that hyperthermia enhances the cell kill; this is evident from the further reduction in tumor cell concentration.
\vspace{-5pt}

\section{Conclusions}
\vspace{-5pt}

We extended our CHR model to incorporate anisotropic transport.
We compared different transport models and observed that the use of both isotropic and prematurely truncated anisotropic transport equations result in underestimation of the overall tumor cell concentration.  
The introduction of radiation results in significant cell kill.
Moreover, introduction of heat enhances radiosensitivity of the tumor cells, which increases the cell kill. 
We plan to extend our model in subsequent papers such that real-world treatments can be addressed.
We also plan to extend this model to address random tumor media.
This will form a major portion of our work in CHR modeling. 
\vspace{-5pt}

\section{Acknowledgments}
\vspace{-5pt}

The authors acknowledge support under award number NRC-HQ-84-15-G-0024 from the Nuclear Regulatory Commission.
The statements, findings, conclusions, and recommendations are those of the authors and do not necessarily reflect the view of the U.S. NRC.
\bibliographystyle{ans}
\bibliography{bibliography}
\end{document}